\documentclass[%
 reprint,
 amsmath,amssymb,
 aps,prl
]{revtex4-2}

\usepackage{booktabs,array}
\usepackage{graphicx}
\usepackage{dcolumn}
\usepackage{bm}
\usepackage{amsmath,mathtools}
\usepackage{pgfplots}
\usepackage{xcolor}
\usepgfplotslibrary{fillbetween}
\usepgfplotslibrary{patchplots}
\usepackage{multirow}

\definecolor{mygreen}{rgb}{0.1, 0.75, 0.1}

\begin{document}


\title{Density Functional Theory Transformed into a One-electron Reduced Density Matrix Functional Theory for the Capture of Static Correlation}

\author{Daniel Gibney}
\author{Jan-Niklas Boyn}%
\author{David A. Mazziotti}%
\email{damazz@uchicago.edu}%
\affiliation{The James Franck Institute and The Department of Chemistry, The University of Chicago, Chicago, Illinois 60637 USA}%
\date{Submitted January 10, 2022}

\begin{abstract}
Density functional theory (DFT), the most widely adopted method in modern computational chemistry, fails to describe accurately the electronic structure of strongly correlated systems.  Here we show that DFT can be formally and practically transformed into a one-electron reduced-density-matrix (1-RDM) functional theory, which can address the limitations of DFT while retaining favorable computational scaling compared to wavefunction-based approaches.  In addition to relaxing the idempotency restriction on the 1-RDM in the kinetic energy term, we add a quadratic 1-RDM-based term to DFT's density-based exchange-correlation functional.  Our approach, which we implement by quadratic semidefinite programming at DFT's computational scaling of $O(r^{3})$, yields substantial improvements over traditional DFT in the description of static correlation in chemical structures and processes such as singlet biradicals and bond dissociations.
\end{abstract}

\maketitle


\textit{Introduction}- Since its formulation by Kohn and Sham~\cite{KSDFT} in the 1960's Density Functional Theory (KS-DFT) has seen widespread adoption within the chemistry community for the prediction and modeling of various chemically relevant properties such as reaction barriers, vibrational modes, ionization potentials, and molecular geometries~\cite{citationstats,ReactionBarriers,vibrations,tddft,ionization,Geometries}. This widespread adoption is owed to DFT's favorable computational scaling compared to wavefunction-based alternatives such as Coupled Cluster (CC) or Complete Active Space (CAS) theories as well as its ease of use as an effectively black-box method~\cite{scaling1,scaling2}. The inexpensive computational scaling of DFT is a direct result of describing the complex exchange and correlation interactions through the use of an efficient energy functional of the electronic density, referred to as the exchange correlation functional. However, while DFT is in principle exact~\cite{Universalfunctional}, the fact that the universal functional is unknown requires the use of approximations, giving us the so-called "functional zoo" of modern DFT functionals~\cite{functionalZoo} and leading to the creation of a Jacobs Ladder classification scheme by Perdew~\cite{Jacobs_ladder}. \\

While functional developments have resulted in improvements to the description of many physical properties such as reaction barriers and ground-state geometries~\cite{dftforproperties,ReactionBarriers,Geometries}, DFT continues to be plagued by three notable failures: i) the self-interaction error, (ii) the charge transfer (or noncovalent interaction) error, and (iii) the static correlation error arising from degenerate or near-degenerate states~\cite{SIE,disp,3errors}, which is rooted in the fact that KS-DFT is inherently based on a single Slater determinant~\cite{frac_dft}. Additionally, it has recently been noted by Medvedev et al. that while the energies obtained with modern functionals have been systematically improving, the fundamental property of DFT, the ground-state electron density, has become less accurate in recent years~\cite{medvedev}. This suggests that new improvements to DFT's functionals are the result of overfitting of the functional form to reproduce experimental results and without the development of a multi-reference formalism DFT will continue inherently to struggle with the description of strongly correlated systems. Recent attempts to rectify this behavior within DFT have involved utilizing the universal functional's known flat plane behavior~\cite{PPLB,SC1,SC2,SC3} as well as the expansion of KS-DFT into the complex plane to allow for solutions beyond what is possible with traditional idempotent DFT~\cite{MHG}. \\

One-electron reduced density matrix functional theory (1-RDMFT) provides a viable path to solve the static correlation problem in KS-DFT while retaining favorable computational scaling compared to wavefunction based alternatives. 1-RDMFT utilizes Gilbert's theorem~\cite{Gilbert} to express the energy of any chemical system as a functional of the 1-electron reduced density matrix (1-RDM). By using the 1-RDM as the fundamental variable, 1-RDMFT is capable of capturing fractional occupation as required for the description of strongly correlated systems~\cite{RDMFT}. Within this area of work, promising approaches for creating 1-RDM functionals involve using the known functional for two-electron atoms and molecules as a starting point~\cite{NOF,2Electron} as well as reconstructions of the connected (or cumulant) part of the 2-electron reduced density matrix in terms of the 1-RDM~\cite{PNOF1,PNOF2,PNOF3,NOF}. These methodologies, while promising for capturing static correlations, are computationally more demanding than DFT. \\

In this work we demonstrate that DFT can be formally and practically transformed into a 1-RDMFT to address DFT's limitations without sacrificing its computational efficiency.  Unlike previously developed 1-RDMFT theories that dispense with DFT's exchange-correlation functional, we convert DFT into a 1-RDM functional theory by (i) relaxing the idempotency restriction on the 1-RDM to recover the full kinetic energy of the electronic system including the contribution from correlation and (ii) adding a quadratic 1-RDM-based term to DFT's density-based exchange-correlation functional.  The second modification is important because it allows us to capture static correlation including fractional orbital occupations in generic atoms and molecules.  We implement the method by a quadratically constrained semidefinite programming algorithm~\cite{SDP1, SDP2, SDP3, SDP4, QSDP, V21, V22} at DFT's computational scaling of $O(r^{3})$.  We apply our 1-RDMFT algorithm to the dissociation of a set of 11 molecules into radicals as well as the Mott metal-to-insulator transition of H$_4$ where the computed 1-RDMs are comparable in accuracy to those from high-level mulireference wavefunction-based methods. \\

\textit{Theory}- While the universal functional of DFT and 1-RDMFT are usually defined with respect to the wave function, we can define them with respect to the 2-RDM as
\begin{eqnarray}
F[\rho] & = & \min_{^{2} D(\rho)}{ {\rm Tr}( {\hat T} \, ^{1} D) + {\rm Tr}( {\hat U} \, ^{2} D) }\,, \\
F[^{1} D] & = & \min_{^{2} D(^{1} D)}{ {\rm Tr}( {\hat U} \, ^{2} D) },
\end{eqnarray}
respectively, where ${\hat T}$ and ${\hat U}$ are the kinetic energy and electron repulsion, $\rho$ is the density, $^{1} D$ is the 1-RDM, $^{2} D(\rho)$ is the $N$-representable 2-RDM constrained to integrate to $\rho$, and $^{2} D(^{1} D)$ is the $N$-representable 2-RDM constrained to integrate to $^{1} D$.  If the 2-RDM is pure-state $N$-representable (representable by at least one $N$-electron wave function)~\cite{Coleman2000, Mazziotti2012b}, we obtain the pure-state functionals of Gilbert~\cite{Gilbert} but if the 2-RDM is ensemble $N$-representable (representable by at least one $N$-electron density matrix)~\cite{Mazziotti2016a} we obtain the ensemble functionals of Valone~\cite{Valone1980}. \\

Exploiting existing functionals of DFT, we define the 1-RDMFT functional as a correction to the DFT functional
\begin{equation}
F[^{1} D] = F[\rho] + C[^{1} D] .
\end{equation}
Written in this novel format, we observe that the correction converts the DFT functional to the 1-RDMFT functional and hence, must accomplish two separate but related adjustments: (1) addition of the full kinetic energy correction to the explicit part of the functional (part of the functional that depends explicitly upon the 1-RDM) and (2) removal of the kinetic energy correction from the universal part of the functional. In recent work we accomplished (1) by expressing the self-consistent-field Kohn-Sham equations as a semidefinite program (SDP)~\cite{SDP_paper}.  A semidefinite program is a special type of optimization in which a linear objective functional of the matrix $M$ is minimized with respect to both linear equalities and the constraint that $M$ is positive semidefinite (nonnegative eigenvalues), $M \succeq 0$~\cite{SDP2, Mazziotti2007c, SDP1, SDP3}. In the SDP formulation the optimization occurs directly with respect to the one particle RDM and the one hole RDM which are restricted to be positive semidefinite. The SDP formulation, which differs from local or complex extensions of DFT~\cite{MHG,Su2018}, relaxes the usual criterion that the 1-RDM be strictly idempotent, allowing the eigenvalues (natural occupation numbers) of the 1-RDM to become fractional and thereby represent bi- or poly-radical character when the molecular orbital energies in the Kohn-Sham Hamiltonian are degenerate. \\

The caveat ``when molecular orbitals in the Kohn-Sham Hamiltonian are degenerate'' introduces a significant restriction that limits applicability of our recent SDP-DFT~\cite{SDP_paper} to static (multi-reference) correlation arising from orbitals that are nearly but not exactly energetically degenerate.  Here we remove this restriction by implementing (2) above: removing the kinetic energy from the universal part of the functional. The term converting DFT to 1-RDMFT must obey several fundamental conditions---it must: (1) vanish in the limit of no correlation, (2) obey particle-hole symmetry, and (3) reward the formation of fractional occupation numbers for molecular orbitals that are degenerate or nearly degenerate in energy.  Taken together, these conditions suggest the following modified Kohn-Sham energy:
\begin{equation}
E_{\rm MKS} = {\rm Tr}[ \left ( {\hat H}_{\rm KS} - ^{1} W \, ^{1} Q \right ) \, ^{1} D ]\,,
\end{equation}
where ${\hat H}_{\rm KS}$ is the conventional Kohn-Sham Hamiltonian, $^{1} Q$ is the 1-hole RDM$ (= ^{1} I - ^{1} D)$, and $W$ is a general positive-semidefinite weight matrix.  Importantly, the weight matrix introduces significant flexibility into the functional form that can be used to optimize the universal functional correction. Physically, the correction contributes a negative energy to the functional as the 1-RDM deviates from idempotency with the orbital occupations becoming fractional. The magnitude of the negative energy contribution is modulated by the selection of the weight matrix. The simplest possible weight matrix is a scalar multiple of the identity matrix, which we represent as $\prescript{1}{}{W_{I}}$. \\

Because this correction is quadratic in the 1-RDM, its addition to the Kohn-Sham energy generates a quadratic semidefinite program.  While quadratic semidefinite programs can be solved directly, we relax it to a regular semidefinite program~\cite{SDP2, Mazziotti2007c, SDP1, SDP3} which can be solved with the scaling of conventional DFT $O(r^{3})$ by the boundary-point SDP algorithm that we developed for solving the optimization problem in the variational 2-RDM method~\cite{SDP2,Schlimgen2016}. Minimization of the quadratic correction can be relaxed to minimization of the trace of a positive semidefinite slack-variable matrix $M$
\begin{equation}
    M = \begin{pmatrix}
    ^{1} F^{}  & ^{1} D^{} \\
    ^{1} D^{} & ^{1} I^{}
    \end{pmatrix} \succeq 0 \,.
\end{equation}
By constructing $M$ in this form, $^{1} F$ is bounded through the determinant as: $ ^{1} F - ^{1} D \, ^{1} D \succeq 0$.  Therefore, the solution of the Kohn-Sham equations can be replaced by a semidefinite program whose energy
\begin{equation}
E_{\rm MKS} = {\rm Tr}(^{1} H_{\rm KS}  \, ^{1} D ) + {\rm Tr}( ^{1} W (^{1} F - ^{1} D) )
\end{equation}
is a functional of the 1-RDM and the slack variable $^{1} F$. \\

By solving the semidefinite program iteratively until convergence of the energy and the 1-RDM, we obtain a general 1-RDM solution of the 1-RDMFT. Importantly, unlike DFT the solution of the 1-RDMFT returns a general, non-idempotent 1-RDM whose fractional occupation numbers provide critical information about the degree of electron correlation in the chemical system. In contrast to other 1-RDMFTs, our theory is constructed as an inexpensive ($O(r^3)$) correction to existing density functional theories. \\



\textit{Results}- The $\prescript{1}{}{W}$ matrix parameters are fitted to reproduce the dissociation curves of N$_2$ obtained from either complete active space self consistent field (CASSCF)~\cite{CASSCF1,CASSCF2} [6,6] or anti-Hermitian contracted Schr\"odinger equation (ACSE)~\cite{ACSE1,ACSE2,ACSE3} calculations in a 6-31G~\cite{631g} basis set. This allows us to evaluate the differences arising from optimizing $\prescript{1}{}{W}$ for primarily static correlation or the total energy of the system to near the FCI limit, in the case of CASSCF or the ACSE, respectively. The ACSE calculations were performed in Maple using the Quantum Chemistry Toolbox ~\cite{Maple,QCP} while the CASSCF results were obtained using PySCF~\cite{PySCF}. B3LYP~\cite{STEPHENS-B3LYP,BECKE-B3LYP} is utilized as the DFT exchange correlation functional for this work.\\

\begin{figure}[]

\centerline{\includegraphics[width=\columnwidth]{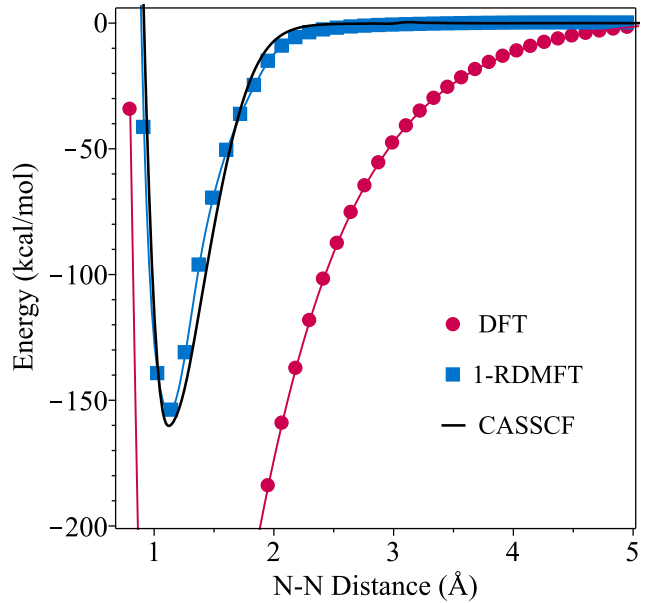}}

\caption{N$_2$ dissociation curves obtained from [6,6]CASSCF, our algorithm with a CASSCF optimized $\prescript{1}{}{W_{I}}$, and B3LYP.}

\label{fig:N2_cas}
\end{figure}

\begin{figure}[h]


\centerline{\includegraphics[width=\columnwidth]{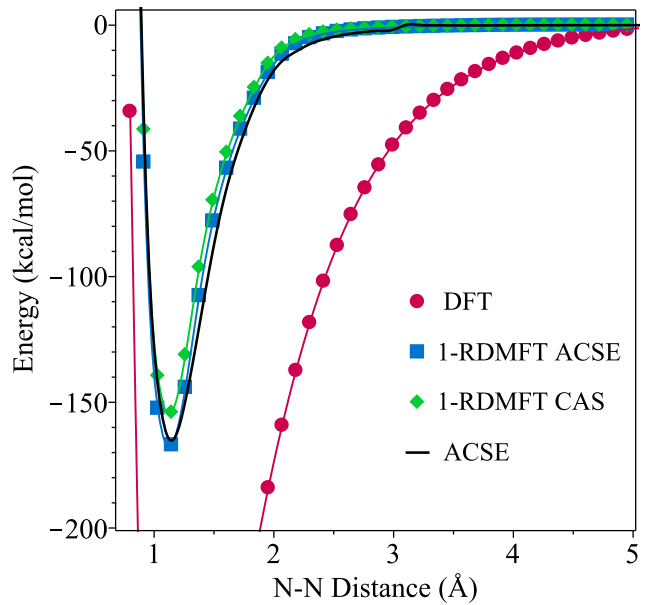}}

\caption{N$_2$ dissociation curves obtained from the ACSE, our algorithm with ACSE and CASSCF optimized $\prescript{1}{}{W_{I}}$ matrices, and B3LYP.}

\label{fig:N2_acse}
\end{figure}

Following optimization of $\prescript{1}{}{W}$ with either CASSCF or ACSE data, we find that the identity-matrix functional form $\prescript{1}{}{W_{I}}$ with a single scalar parameter accurately reproduces the dissociation curve of N$_2$ with only minor deviations. The dissociation curve obtained from the $\prescript{1}{}{W_{I}}$ functional form trained on CASSCF is compared to CASSCF and B3LYP in Fig.~\ref{fig:N2_cas}. Use of the 1-RDM correction yields major improvements over traditional B3LYP, closely mirroring the CASSCF curve and recovering the correct behavior in the dissociated regime. The CASSCF dissociation energy is recovered to within 5.53 kcal/mol with the $\prescript{1}{}{W_{I}}$ correction. Minor deviations from CASSCF arise in the intermediate bonding region where dynamical correlation plays a major role. Training of the $\prescript{1}{}{W_{I}}$ correction on the ACSE solution further improves the description of the bonding and intermediate bonding regions and the ACSE dissociation energy is again reproduced with high accuracy, yielding a deviation of 1.69 kcal/mol for the $\prescript{1}{}{W_{I}}$ correction. The two $\prescript{1}{}{W_{I}}$ functionals trained on the ACSE and CASSCF are displayed with the ACSE curve in Fig.~\ref{fig:N2_acse}.  \\


\begin{figure}[]


\centerline{\includegraphics[width=\columnwidth]{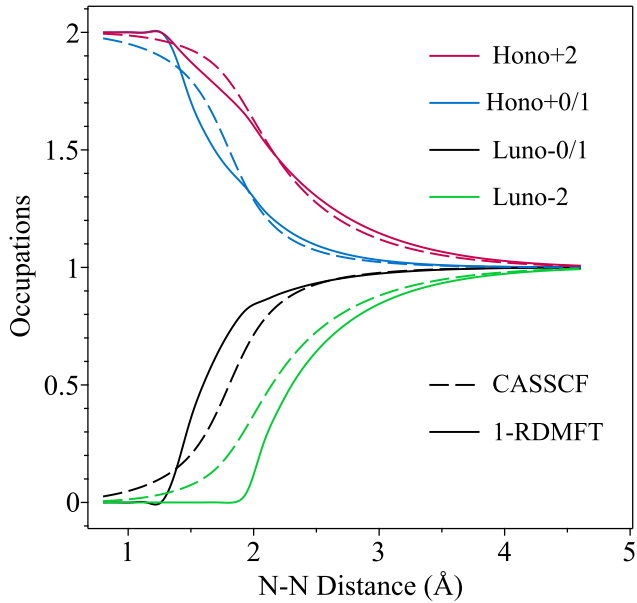}}

\caption{Orbital occupations obtained for the 6 frontier orbitals from a CASSCF optimized $\prescript{1}{}{W_I}$ algorithm (solid lines) and [6,6] CASSCF (dashed lines) for the dissocation of N$_2$ in a 6-31G basis set.}

\label{fig:N2_occ}
\end{figure}

Investigation of orbital occupations along the dissociation coordinate (displayed in Fig.~\ref{fig:N2_occ}) reveals that our algorithm, while optimized to reproduce the energy along the dissociation curve, also recovers orbital occupations in line with those of the CASSCF [6,6] calculation. In the bonding region of 1~{\AA} there are no deviations from integer occupations and single-reference behavior is recovered. However, as the bond is stretched the six strongly correlated frontier orbitals deviate from their integer occupations with the deviations approaching half occupation at 4~{\AA}, while the remaining orbitals display the correct integer occupations. \\

Using the $\prescript{1}{}{W_{I}}$ matrices fitted to the N$_2$ dissociation, we apply our algorithm to a set of 11~molecules, B$_2$, C$_2$, CN, CO$_2$, CO, F$_2$, N$_2$, NF$_3$, NO, S$_2$, SiO, selected from the MR-MGN-BE17 test set~\cite{MN19DB}, which is designed to benchmark a method's ability to describe multi-reference correlation in bond dissociation. Results obtained with $\prescript{1}{}{W_{I}}$ fit to the CASSCF and ACSE dissociation curves of N$_2$ are compared to the respective CASSCF and ACSE reference data. In order to maintain consistency between the different molecular systems, the CASSCF active space was constructed to include those molecular orbitals that display a significant contribution from the valence atomic p-orbitals.\\

We first consider the dissociation energies for the CASSCF optimized $\prescript{1}{}{W_{I}}$ and data are shown in the top rows of Table \ref{tab:dissociations}. It is evident that B3LYP substantially overestimates the dissociation energies as it fails to capture the static correlation of the dissociated systems, yielding a mean unsigned error (MUE) of 154.09 kcal/mol. In contrast, the 1-RDMFT displays a MUE of 32.19 kcal/mol for the adjustment $\prescript{1}{}{W_I}$, yielding a five fold reduction in error compared to traditional KS-DFT. Additionally, our algorithm results in a sign change of the mean signed error (MSE), on average underestimating the dissociation energy whereas KS-DFT consistently overestimates it. \\

\begin{table}[]
    \centering
    \caption{Errors of the dissociation energies in kcal/mol obtained with CASSCF fitted $\prescript{1}{}{W_{I}}$ functionals and B3LYP as compared to the [6,6] CASSCF and ACSE references. }
    \begin{ruledtabular}
    \begin{tabular}[b]{cccc}
          \multicolumn{4}{c}{CASSCF OPTIMIZED} \\
            REF & & B3LYP & $\prescript{1}{}{W_I}$\\
      \hline
     \multirow{2}{*}{CASSCF} & MSE    & 154.09 & -17.68\\
     & MUE    & 154.09 & 32.19\\
      \hline
     \multirow{2}{*}{ACSE} & MSE    & 137.03 & -34.73\\
     & MUE    & 137.03 &  39.83\\
     \hline
    \multicolumn{4}{c}{ACSE OPTIMIZED} \\
    \hline
     \multirow{2}{*}{ACSE} & MSE & 137.03 & -27.76\\
     & MUE & 137.03 & 36.19\\
    \end{tabular}
    \end{ruledtabular}
    \label{tab:dissociations}
\end{table}

The calculations are repeated with the adjustment optimizations based on the dissociation curve of N$_2$ obtained with the ACSE. Recovery of additional dynamical correlation in the training data improves the results obtained with our 1-RDM functional corrections. The ACSE reference columns of Table~\ref{tab:dissociations}) compare the CASSCF and ACSE trained functional forms. ACSE optimization reduces the $\prescript{1}{}{W_I}$ adjustments MUE's by approximately -3 kcal/mol. Errors for the individual molecular systems in the data set can be found in the Supplementary Information. Overall our algorithm's adjustments again yield significantly lower errors in the dissociation energy of the small molecules in the set surveyed compared to traditional B3LYP. \\

\begin{figure}[h]

\centerline{\includegraphics[width=\columnwidth]{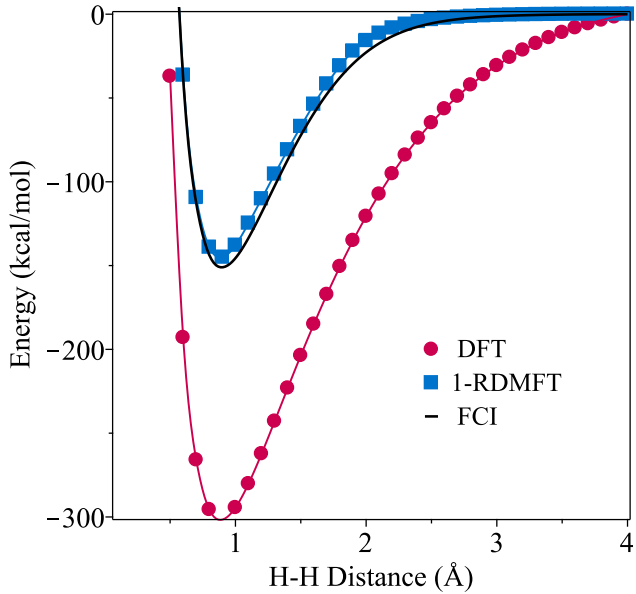}}

\caption{Symmetric dissociation of linear H$_4$ in a 6-31G basis from B3LYP, FCI, and our algorithm with an CASSCF optimized $\prescript{1}{}{W_{I}}$.}

\label{fig:H4}
\end{figure}

\begin{figure}[h]
\centering

\centerline{\includegraphics[width=\columnwidth]{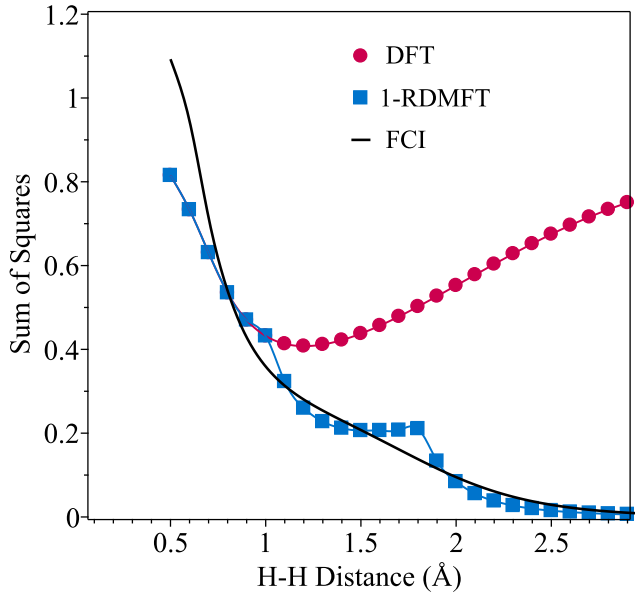}}

\caption{Sum of squares of the off diagonal elements of the 1-RDM obtained as a function of interatomic distance for linear H$_4$. The sum is defined as $\sum_{i\neq j}{{D^i_j}^{2}}$ where i and j are atomic centers.}

\label{fig:Sum_Squares}
\end{figure}

Lastly, we investigate the Mott metal-to-insulator transition of linear H$_4$ using our algorithm with the $\prescript{1}{}{W_{I}}$ adjustment, and compare the results to those obtained from FCI~\cite{FCI1,FCI2} and B3LYP using a 6-31G basis set~\cite{631g}. Figure~\ref{fig:H4} displays the dissociation curves obtained with the $\prescript{1}{}{W_{I}}$ adjustment, KS-DFT with the B3LYP functional, and FCI. The $\prescript{1}{}{W_{I}}$ based functional adjustment yields a dissociation error of just 5.9~kcal/mol when compared to FCI. Additionally, the functional adjustment is able to prevent the convergence failure seen in DFT at large interatomic distances, yielding smooth asymptotic behavior in the dissociation limit.\\

As the linear hydrogen chain dissociates, it should become an insulator with the elements of the 1-RDM in the local atomic-orbital basis set corresponding to different atoms decaying to zero. The sum of these 1-RDM elements squared in Fig.~\ref{fig:Sum_Squares} correctly tends towards zero as the interatomic distance is increased for the $\prescript{1}{}{W_{I}}$ correction, recovering FCI behavior, while in B3LYP it erroneously approaches 0.83, indicating that in DFT the hydrogen chain remains metallic.  This failure in traditional KS-DFT arises from the fact that KS-DFT, being implemented in a non-interacting limit, imposes idempotency on the 1-RDM, which prevents the electron correlation that is necessary to localize the density in the dissociation limit.

\textit{Conclusions}
Density functional theory (DFT), we show, can be formally and practically transformed into a one-electron reduced-density-matrix functional theory (1-RDMFT), in which the exact correlated 1-RDM is used as the basic variable in the theory.  Use of the exact 1-RDM allows us to address the limitations of DFT in its treatment of static correlation that often leads to inaccuracies in the prediction of molecular properties, especially at non-equilibrium geometries.  In contrast to most 1-RDMFTs, we retain the use of DFT's exchange-correlation functional of the density.  While recent work including our own has relaxed the idempotency restriction on the 1-RDM in the kinetic energy term, here we also add a quadratic 1-RDM-based term to DFT's density-based exchange-correlation functional.  This additional term is especially important because it allows us to treat static correlation in molecular systems without energetically degenerate frontier orbitals.  The retention of DFT's exchange-correlation functional allows us to implement the 1-RDMFT algorithm by quadratic semidefinite programming at DFT's computational scaling of $O(r^{3})$.  The 1-RDMFT, as shown in the results, yields significant improvements over traditional DFT in the description of static correlation in chemical structures and processes such as singlet biradicals and bond dissociations.

Accurate descriptions of molecular bond dissociations are achieved with a simple form of $\prescript{1}{}{W}$ based solely on the identity matrix modified by a scalar term. Using only N$_2$ as the training set this functional form displays significant improvements over traditional KS-DFT in the capture of the dissociation energies of small molecules. Furthermore, while our parameter training focused solely on the capture of static correlation within the $\pi$ and $\pi^*$ orbitals of N$_2$, the resulting 1-RDM functionals accurately described the Mott insulator transition of H$_4$, a system lacking statically correlated $\pi$ orbitals.  Moreover, even though functional optimization for the capture of fractional natural occupation numbers is not performed, the 1-RDM correction in all systems studied yields natural occupations in line with those obtained from CASSCF.\\

In summary, the present work provides a foundation for further research into the capture of multi-reference correlation through the inclusion of fractional 1-RDMs in a DFT framework. The functional developed in this work, which is related to the recent addition of a correction to the Hartree-Fock energy~\cite{Wang2022}, shows significant promise for the development of an efficient 1-RDM functional theory that uses existing DFT exchange-correlation functionals, or modifications thereof, for the evaluation of electronic energies of strongly correlated systems.  \\

\textit{Acknowledgements}

D.A.M. gratefully acknowledges the U.S. National Science Foundation Grant No. CHE-1565638.

\bibliography{citations.bib}

\end{document}